\documentclass[conference]{IEEEtran}
\newcommand{\spacetune}[1]{#1}
\newcommand{\fname}[1]{\texttt{"#1"}}
\newcommand{\code}[1]{{\texttt{#1}}}
\newcommand{\reduce}{{\sc reduce}}

\usepackage{graphicx}
\usepackage{comment}
\usepackage{balance}
\usepackage{xcolor}
\definecolor{linkc}{rgb}{0.1,0.1,.8}
\definecolor{darkgreen}{rgb}{0,0.5,0}
\definecolor{midblue}{rgb}{0,0,0.7}
\usepackage{url}
\usepackage[bookmarks=false,colorlinks=true,citecolor=darkgreen,linkcolor=linkc,filecolor=linkc,urlcolor=linkc]{hyperref}

\usepackage{amsmath}\interdisplaylinepenalty=2500
\usepackage{algorithmic}
\usepackage[caption=false,font=footnotesize]{subfig}
\usepackage{cleveref}
\usepackage{cite}      % Better citations
\usepackage{array}     % Better array spacing
\begin{document}

\author{
\IEEEauthorblockN{Arthur C. Norman}
\IEEEauthorblockA{Trinity College\\
Cambridge, UK \\
\texttt{acn1@cam.ac.uk}}
\and
\IEEEauthorblockN{Stephen M. Watt}
\IEEEauthorblockA{Cheriton School of Computer Science, U. Waterloo \\
Waterloo, Canada\\
\texttt{smwatt@uwaterloo.ca}}
}

\title{Semi-Centennial REDUCE}
\date{}
\maketitle
\begin{abstract}
We present a version of the \reduce{} computer algebra system as it was in the early 1970s.
We show how this historical version of \reduce{} may be built and run in very modest present-day environments and outline some of its capabilities.
This shows that a large part of computer algebra function can be delivered by a system that is small by today's standards and can be deployed in resource-constrained settings.
\end{abstract}
\maketitle
\pagestyle{empty}

\spacetune{\setlength{\parskip}{1.1ex}}
%%%%%%%%%%%%%%%% Actual Text %%%%%%%%%%%%%%%%%%%%%%%%%%%%%%%%%%%%%%
~\\
\section{Introduction}

The \reduce{} computer algebra system has its origins in the late 1960s, growing out of a set of specialized programs for symbolic computation in high energy physics.   
By the early 1970s, these had evolved to a form similar to what we see as \reduce{} today and already providing many of the essential capabilities of modern computer algebra systems.

We note that even the most meagre of today's common computing platforms are significantly more powerful than the hardware on which this version of \reduce{} was deployed.   
One may therefore contemplate how significant computer algebra code can be incorporated in small devices such as micro-controllers.  This takes us further into the territory explored by the \textit{Compact Computer Algebra} workshop series, \textit{e.g.}~\cite{Fitch09}.

Both authors have a historical connection with \reduce{}, though to greatly different degrees.
One of us (ACN) has been one of the principal contributors to the system over almost its entire existence, \textit{e.g.}~\cite{Norman93}.  The other of us (SMW) was a user of  \reduce{} in the 1970s for general relativity computations:  after performing the first of two similar lengthy computations by hand over the course of three months (as required by a thesis advisor), it was possible to learn \reduce{} and complete the second computation in one weekend~\cite{watt81}. 

In what follows, we summarize some of the key steps that led to this early version of \reduce{}, outline the issues encountered in building a working version from archival sources, summarize some of its capabilities and discuss its performance.
This 1973 version of the software is publicly available for download~\cite{Reduce1973Download}.
The software portability issues encountered in its development are recounted in~\cite{norman-watt:ca-portability}
and the historical context is given in~\cite{Corless+:25}.

\section{Early REDUCE}
In 2005 Tony Hearn, the originator of the \reduce{} algebra system, presented a paper ``\reduce{} --- The First Forty Years''~\cite{Hearn:05}, explaining that in the 1960s he was working on calculations involving Feynman Diagrams and that John McCarthy suggested he have a go at using the new language Lisp to mechanize some of the really tedious work, and also offered him access to an IBM 7090. At the time, this was a state of the art large computer suitable for such a project. 
By the start of the 1970s, the code that had originated as specifically aimed at High Energy Physics calculations had extended to become suitable for a range of uses. 
In 1971 he presented this work
to a fully international audience~\cite{Hearn:71a}.
Some important papers leading up to this
were~\cite{hearn66,hearn68,hearn69}.
Building on collaboration with people he met at that and other conferences, he prepared a test and demonstration script for \reduce{} that included samples of calculations relevant to celestial mechanics and general relativity as well as high energy physics and rather general basic computation. In 1985 Marti and Hearn collected
timings for a wide range of machines and reported
these in ``\reduce{} as a Lisp Benchmark''~\cite{Marti:85}.
A version of that test script is part of the set of regression tests shipped with the system in its open source form even today.

Fairly recently some archive tapes from the early 1970s were recovered and decoded, and found to contain a set of \reduce{} sources from 1972--73. Inspection shows that the user interface they offer has hardly changed in the intervening 50 years and that quite large segments of the code are completely recognizable, with only a conversion from all upper case to lower case lettering. The version of the test
script included there differs in a very few rather minor ways from the one used today. It was possible to build and run this historic version of the software.

The version we have recovered and made available for demonstration now sits on a naively-implemented and fairly minimal Lisp system coded in C that uses just an interpreter (\textit{i.e.} no compiler) and so by today's standards the performance there is severely compromised. 
However, on even cheap machines today such as the Raspberry Pi, we can
compare its absolute timings with a wide range of systems from the past that were reported in the benchmark paper previously cited. The times we see are similar to those on mainframes from earlier times and with specialized workstations such as
dedicated Lisp machines and the amount of store we can now deploy is absurdly greater than any of the early-days options.

By reviewing this version of the system from the early 1970s we can see the growth in capability (and bulk of source code) that development with multiple international contributors has grown \reduce{} to, but also be reminded how many of the essential core capabilities that it offers were available that far back. And we can be reminded of the impact of Moore's Law and realize just what a challenge getting anything going well would
have been that long ago.

\section{Building The 1973 System Today}
\label{building}
The files we have used are available for download~\cite{Reduce1973Download}.
They derive mostly from a version of the \reduce{} source from 1973 as recovered from some archive files.
%The comments here are by ACN reporting on the experience of getting the code to run.
%
This version was for TENEX on a PDP 10 but we also
have sources dated September 1972 where the comments make it clear that it
was for use on the MTS system running on an IBM 360. The PDP10 version was
slightly more complete and easier to work from.

The main \reduce{} source code is found in the file \fname{reduce.red}.
The process of making this
run will have been very similar to the porting process normally needed
for \reduce{} then when Lisp systems on the various machines it was used on
differed. The start of the file contains a collection of mappings,
definitions and redefinitions that Lisp 1.6 on TENEX did not support
or where it used different names. 
It was necessary to replace or override these
with code that has the same intent but made compatible with the
small Lisp system  used.

The code shows signs of a range of compatibility issues it had encountered.
The ones that seem most notable 
are enumerated below.
The PDP10 scheme for covering all of this is just a clearly
separated out 500 line prelude to the main \reduce{} code.
%As in item (\ref{itemone}) above, it is clear that LISP/360 was not offering a string data type. 

%
\begin{enumerate}
\setlength\itemsep{4pt}
\item
\label{itemone}
 Some Lisp systems supported strings (as in \code{"hello"}) as
    a native data structure while other did not. It looks as if \reduce{}
    could at one time run using a representation
    
    \code{(STRING ORDINARYSYMBOL)}.
\item
 There are functions \code{EXPLODE} and \code{COMPRESS} that map between symbols and numbers (and perhaps strings) and lists of characters. The behaviour of these on symbols whose names included whitespace or punctuation marks was not fully stable. Specifically using \code{COMPRESS} on a list consisting of \code{(* A B C)} might produce a symbol called \code{*ABC} or it might stop at the ``\code{*\/}'' and need escape characters to support unusually spelt characters. And then \code{EXPLODE} might or might not insert those escape characters.
\item
 The Lisp might or might not provide vectors or arrays, and if it did
    the names of the functions involved were not standardized. \reduce{} could potentially exploit a rich Lisp, but otherwise would represent a vector as a list. If the underlying Lisp provided a function \code{VECTORP} that tested for a ``real'' vector but \reduce{} was using its emulated ones confusion could arise.
\item
 If \code{FN} is the name of a Lisp variable, should \code{(FN ARG)} invoke the function or would it be necessary to use \code{(APPLY FN (LIST ARG))}? The terminology used regarding this was ``Is it a Lisp-1 or a Lisp-2?''
 This issue persists today between Scheme and Common-Lisp variants.
\item
 A whole range of small utility functions might either not be provided by the Lisp or might be present but under a different name --- or with arguments taken in a different order!
\item
\label{itemsix}
 Back in the early 1970s taking the stream of input characters and turning it into a stream of symbols felt expensive, and any particular Lisp might have desirable built-in functions that could really help but would not port to other platforms.
\item
 Character classification in support of item \ref{itemsix} could be untidy. See for instance this fragment from the file \fname{reduce.red} that identifies letters:
 
% For \usepackage{fancyvrb},
%\begin{Verbatim}[fontsize=\small,fontfamily=\rmfamily,shape=slanted] 
\begin{small}
\begin{verbatim}
SYMBOLIC PROCEDURE LITER X;
   NULL NUMBERP X AND
   (X := LSH (MAKNUM(CAAR GET(X,'PNAME),
                     'FIXNUM),-11))>64
    AND 91>X;
\end{verbatim}
\end{small}
%\end{Verbatim}

\noindent
    where the \code{MAKNUM} and the shift are not very portable and it is clear that lower case letters were not been considered.
\item
 The file \fname{reduce.red} contains a 4-line assembly code function for use for ordering variables. That needs adjustment.
\item
 Anything that is specific to interactive \textit{vs} batch use is of course ``delicate''.
\end{enumerate}

Once \reduce{} is built it must be able to parse its own source --- but of
course at the start of testing it has not been built! So to bootstrap
it one starts by using an existing running version of \reduce{} to
parse the file \fname{reduce.red} and emit the raw Lisp equivalent. 
In the files, there is also a \reduce{} source file dated 15 September 1972  that is all in Lisp. 
It includes many checks of the form  \code{(EQ *MODE (QUOTE SYMBOLIC))} and
that amounts to a signature of a system where executable parts of \reduce{} would be written in its own language. 
A version of \reduce{} in its
own language needed a part directly Lisp-coded for bootstrapping, so Lisp
versions like this will have persisted for a while even when the main
development had moved on. 

For pragmatic reasons some of the machine-sensitive customizations are done
ahead of that conversion. A current 
(\textit{i.e.} 2025) 
version of \reduce{} was used
to make a file \fname{reduce.lsp}. For this conversion a number of behaviours
that the current parser would have exhibited needed to be disabled, either
by removing tags from property lists or by hiding some symbols completely.
The file \fname{reduce.red} was then hand-edited further to gain the level of
compatibility between 1973 and 2025 standards that was needed.

The resulting \fname{reduce.lsp} was loaded into a Lisp system known as "vsl".
This is a fairly minimal Lisp that consists of 3500 lines of C code
implementing much of the "Standard Lisp" that later versions of \reduce{}
settled on. It is complete enough that it can run an up to date version
of \reduce{}, but is an interpreter-only system with no optimizations and so is not at all fast.
The aim was that the C code should be fairly spartan, and so for instance
at that level it supports only \code{CAR} and \code{CDR}. All the combinations such as
\code{CDADR} and so on are then defined in Lisp in the file \fname{vsl.lsp}. That file
also provides the functions to append and reverse and search lists. But the
two big sections of code in \fname{vsl.lsp} are the ones that support big integers
and that can prettyprint (\textit{i.e.} print with neat indentation) Lisp code.
In 1973 many Lisp systems did not support integers with more bits than
would fit in a machine register. The file \fname{vsl.c} provides for a representation of
overlarge integers but leaves all arithmetic on them to lisp code in \fname{vsl.lsp},
where the numbers are handled as list of digits. Performance took second
place to code simplicity.
It would have been possible to modify
vsl to support the ancient \reduce{} more directly,
but instead 
\fname{reduce.lsp} was modified to live with it.
We note that \fname{vsl.c} uses \texttt{zlib} compression so that saved heap images consume less space and  \texttt{histedit} to support more civilised terminal interface.  Neither of these is intrinsically necessary for \reduce{} -- they are just used in the small Lisp system to make life easier.

\spacetune{\enlargethispage{\baselineskip}}
The result of all of the above is a workable \reduce{} (which will without
doubt have residual errors and limitations above and beyond the ones
inherent from its age). The test applied for it is a file that in
a modern copy of \reduce{} would live as packages/alg/alg.tst and which
was for very many years treated as the definitive demonstration and main
test for \reduce{}. Tony Hearn collected timings from running it on many
computers (dedicated Lisp machines, departmental servers, mainframes,
supercomputers and various more personal machines) and today it feels
amazing that the really crude implementation reported here is dramatically
faster then a whole range of impressive and expensive hardware from the
past~\cite{Marti:85}. 
\begin{comment}
As what may be a key illustration, 
this test ran for most of a minute on a VAX 11/780 or for 5 or 6 seconds
on the fastest IBM mainframe of the day. The current crudely implemented
interpreter-only Lisp running on a 7 year old (and hence not really
cutting edge) x86\_64 PC completes the run in about 0.3 second.
\end{comment}

Note that the underlying representation used for
polynomials and for rational functions has importantly not changed
from usage in the early sources. The functions \code{ADDF}, \code{ADDSQ} and flags such
as \code{!*MCD}, along with the general framework for simplifications using a
function \code{SIMP} that dispatches based on the identity of the expression it
is handed has the same overall structure even though a range of detailed
changes and upgrades have been applied. The same holds true for much of the code for both parsing and printing: the function names in the old code
will be really familiar to those who work with the current system.

\section{The System Capabilities in Brief}
This 1973 version of \reduce{} was fully capable of polynomial and rational function arithmetic and computations involving elementary and unknown functions, including user-defined simplification rules.

Reverting a few minor syntax changes, described below, the current standard \reduce{} test file, \fname{alg.tst}, is able to run in the 1973 system.  That file shows:
\begin{itemize}
    \item programmable control flow, variables, arrays, and user defined procedures,
    \item solution of the $f$ and $g$ series problem~\cite{Sconzo65},
    \item a problem in Fourier analysis,
    \item a general relativity program to compute Christoffel symbols and the Riemann, Ricci and Einstein tensors from any given metric tensor,
    \item solution to a problem from the textbook \textit{Relativistic Quantum Mechanics}~\cite{Bjorken+64}.
\end{itemize}
The contents of the test file  for the 1973 \reduce{}  is  given in the appendix.
To be compatible with the 1973 version of \reduce{}, it was necessary to modify the current standard test file \fname{alg.tst} in~\cite{Reduce1973Download} to older syntax as follows:
\begin{itemize}
\item
replace \small \code{<<...>>} blocks with \code{BEGIN ... END},
\item unwind the use of \code{WHERE} into \code{LET}, and
\item replace the result variable \code{ws} with  \code{!*ans}.
\end{itemize}

Despite its early appearance and small size, reviewing various test and demonstration files
that came with \reduce{} and the reports on applications in published work
citing it, suggests that it could address a broad swathe of problems. Users
would often need to re-cast their problems to allow for its restrictions
(\textit{e.g.} notably avoiding need for the calculation of polynomial GCDs
wherever possible, for instance by introducing new symbols to stand for
more complicated expressions that might appear in denominators). And where
there were intrinsic limitations (such as a lack of floating point
arithmetic or support for big integers) those were in fact addressed
over the next few years.

\spacetune{\enlargethispage{-\baselineskip}}
\section{Observations}
We can compare the 1973 version of \reduce{} and its computing environment to those of today.
The source is a little over 6000 lines long
(170 Kbytes) as against the current version which is over 500,000 lines
and almost 19 Mbytes.
Note that Reduce 3.8, a version from 2007 fairly shortly before the code
was released as open source, was 300K lines and 10 Mbytes.

In terms of performance, we must understand the vast improvement of typical computing power that has occurred.
In 1973, the IBM 370-168 model 1 was a \textit{high-end} modern machine on which one might run \reduce{}.  Today, it is hard to find a \textit{lower-end} general purpose computer than a Raspberry Pi Zero 2W.  These compare as shown in Table~\ref{ibm-vs-rpi}.
While this gives only a very rough comparison, we see that the cost of computing, measured as MHz per 2024 {\sc usd}, has improved by a factor of about 3 or 4$\times 10^9$.
Even for the very limited modern Raspberry Pi Zero platform, the amount of primary memory is 64 times that of the high-end machines of 1973.   Comparing the IBM 360-168 mainframe to a high end modern laptop (MacBook Pro 2023), the available primary memory can now be 256,000 times greater and secondary memory 40,000 times greater.
\begin{table}[t]
    \centering
    \begin{tabular}{lcc}
         & \textbf{IBM}             & \textbf{Raspberry} \\
         & \textbf{370-168 model 1} & \textbf{Pi Zero 2W} \\
    \hline
    \\[-2.5mm]
    Year Introduced     &  1972          &  2021 \\
    Cost ({\sc usd} 2024) & 35--50,000,000 &  15  \\
    Size (m$^2$)          & 90-140         & 0.002 \\
    Power w/o periph. (W) & 50--70,000     & 0.6 \\
    Cooling               & active water   & passive air\\
    Instruction Set       & CISC           & RISC \\
    Clock Speed (MHz)     & 2              & 2400 \\
    Word Size (bits)      & 32             & 64 \\
    Primary Mem. (MB)     & 1--8           & 512 \\
    Secondary Mem. (MB)   & 200            & 64,000 (typical)\\
    Video    & 24$\times$80 mono CRT & 1080p30\\
    \hline
    \end{tabular}\\
    ~\\
    \caption{High-end 1973 computer vs low-end 2025 computer}
    \label{ibm-vs-rpi}
    \spacetune{\vspace{3\baselineskip}}
\end{table}

With this in mind, we can compare the time to run the \fname{alg.tst} test file on various historical computer models versus the time to run it with the 1973 version of \reduce{} on modern hardware.  A few timings are shown in Table~\ref{timing-comparison}, where some notable historical times are taken from~\cite{Marti:85}.
The  DEC KL-10 was the machine on which {\sc macsyma} was made available over the ARPANET,  the VAX 11/780 was the
machine on which most of {\sc Maple}'s initial development took place,
the IBM 308X series was the platform on which the {\sc scratchpad ii} (later named {\sc axiom}) had most of its initial development, and the Symbolics 3600 was a single-user computer optimized for Lisp and featuring {\sc macsyma}.
The Xerox Dolphin is representative of the early years
and the largest memory of these historical machines reported was the Symbolics 3600, which could be configured with up to 7.5 megawords of memory (around 32 Mbytes).  The VAX 11/780 and IBM 3084 were the mainstays of academic use in their time.

We demonstrated the 1973 version of \reduce{} running on a Raspberry Pi
Zero at ISSAC 2025. Rather than using a modern, optimised, full-scale Lisp
system in this demonstration we chose to build Reduce on top the na\"{\i}ve ``vsl'' Lisp system as described earlier,  and whose C source
is included with the historical \reduce{}. 
%This little Lisp is in
%fact more capable that is necessary for the support of the 1973 code in
%that a variant of it can run the fully up to date version of \reduce{},
%albeit somewhat slowly. 

When we compare \reduce{} speed
on the cut-down Lisp 
%(that the timings reported here came from) 
with a version running on a production-quality Lisp, we see a speed factor of
about 6, so current computers, even tiny ones, are even faster as
compared with the biggest ones of the past than the figures we have
presented might initially suggest.

\spacetune{\newpage}
\section{Conclusions}
We have shown that a version of \reduce{} from more than 50 years ago can be built and run today, and that it was already highly capable.
We have detailed the process of getting it running on a minimal Lisp and have seen that doing so recalls all the Lisp portability issues from that era. 
The resulting system is tiny by today's standards
and requires no more resources than those of current microcontrollers. % see ulisp.com

\begin{table}[t]
    \centering
%% Timings from Marti and Hearn 85
\begin{tabular}{lr}
\textbf{Model}        & \textbf{CPU time (s)} \\
\hline
    \\[-2.5mm]
    \multicolumn2l{
\textit{Contemporary \reduce{} versions on then 
state-of-the-art Lisps} \hfill}\\[.1mm]
Xerox Dolphin         & 322.0       \\
DEC KL-10             &  24.6       \\
DEC VAX 11/780        &  50.3       \\
Symbolics 3600        &  45.0       \\
IBM 370 158           &  49.9       \\
IBM 3084              &   5.4       \\
\hline
    \\[-2.5mm]
    \multicolumn2l{
\textit{1973 \reduce{} on the na\"{\i}ve Lisp described in Sec~\ref{building}\hfill}}
\\[.1mm]
Raspberry Pi Zero 2W  & 4.416       \\
Raspberry Pi 5        & 0.599       \\
MacBook Pro (M3 Max)  & 0.239       \\
\hline
    \\[-2.5mm]
    \multicolumn2l{
\textit{Modern \reduce{} on Cambridge Standard Lisp} \hfill}
\\[.1mm]
Raspberry Pi 5        & 0.099       \\
MacBook Pro (M3 Max)  & 0.050       \\
\hline
\end{tabular}\\
~\\
    \caption{Time to run \fname{alg.tst}.  Lower is better.
    }
    \label{timing-comparison}
    \spacetune{\vspace{-1.4\baselineskip}}
\end{table}
\bibliographystyle{plain}
\IfFileExists{IfExistsUseBBL.tex}{%

}{%
\bibliography{main.bib}
}
\newpage
\onecolumn
\section*{Appendix: Input file \fname{alg.tst}}
\begin{footnotesize}
\begin{verbatim}
(begin)
COMMENT This is a standard test file for REDUCE that has been used for
many years.  It only tests a limited number of facilities in the
current system.  In particular, it does not test floating point
arithmetic, or any of the more advanced packages that have been made
available since REDUCE 3.0 was released.  It does however test more
than just the alg package in which it is now stored.  It has been used
for a long time to benchmark the performance of REDUCE.  A description
of this benchmarking with statistics for REDUCE 3.2 was reported in Jed
B. Marti and Anthony C. Hearn, "REDUCE as a Lisp Benchmark", SIGSAM
Bull. 19 (1985) 8-16.  That paper also gives information on the the
parts of the system exercised by the test file.  Updated statistics may
be found in the "timings" file in the REDUCE Network Library;

COMMENT to be compatible with the 1973 version of Reduce I have needed
to remove use of "<< ... >>" block and replace them with "BEGIN ... END",
unwind use of "WHERE" into "LET" and replace "ws" with "!*ans". I believe all
these are rather simple changes not representing deep adjustment to the
system capabilities;
showtime;

COMMENT some examples of the FOR statement;

COMMENT summing the squares of the even positive integers through 50;

for i:=2 step 2 until 50 sum i**2;

COMMENT to set  w  to the factorial of 10;

w := for i:=1:10 product i;

COMMENT alternatively, we could set the elements a(i) of the
        array  a  to the factorial of i by the statements;

array a(10);

a(0):=1$

for i:=1:10 do a(i):=i*a(i-1);

COMMENT the above version of the FOR statement does not return
        an algebraic value, but we can now use these array
        elements as factorials in expressions, e. g.;

1+a(5);

COMMENT we could have printed the values of each a(i)
        as they were computed by writing the FOR statement as;

for i:=1:10 do write a(i):= i*a(i-1);

COMMENT another way to use factorials would be to introduce an
operator FAC by an integer procedure as follows;

integer procedure fac (n);
   begin integer m;
        m:=1;
    l1: if n=0 then return m;
        m:=m*n;
        n:=n-1;
        go to l1
   end;

COMMENT we can now use  fac  as an operator in expressions, e. g.;

z**2+fac(4)-2*fac 2*y;

COMMENT note in the above example that the parentheses around
the arguments of FAC may be omitted since it is a unary operator;

COMMENT the following examples illustrate the solution of some
        complete problems;

COMMENT the f and g series (ref  Sconzo, P., Leschack, A. R. and
         Tobey, R. G., Astronomical Journal, Vol 70 (May 1965);

deps:= -sigma*(mu+2*epsilon)$
dmu:= -3*mu*sigma$
dsig:= epsilon-2*sigma**2$
f1:= 1$
g1:= 0$
for i:= 1:8 do 
 begin f2:=-mu*g1 + deps*df(f1,epsilon) + dmu*df(f1,mu) + dsig*df(f1,sigma);
   write "F(",i,") := ",f2;
   g2:= f1 + deps*df(g1,epsilon) + dmu*df(g1,mu) + dsig*df(g1,sigma);
   write "G(",i,") := ",g2;
   f1:=f2;
   g1:=g2 end;

COMMENT a problem in Fourier analysis;

for all x, y let cos(x)*cos(y) = (cos(x+y)+cos(x-y))/2,
               cos(x)*sin(y) = (sin(x+y)-sin(x-y))/2,
               sin(x)*sin(y) = (cos(x-y)-cos(x+y))/2,
               cos(x)**2 = (1+cos(2*x))/2,
               sin(x)**2 = (1-cos(2*x))/2};


factor cos,sin;

on list;

(a1*cos(omega*t) + a3*cos(3*omega*t) + b1*sin(omega*t)
		 + b3*sin(3*omega*t))**3;

remfac cos,sin;

off list;

for all x,y clear cos(x)*cos(y),cos(x)*sin(y),sin(x)*sin(y),
                  cos(x)^2, sin(x)^2;
COMMENT Leaving such replacements active would slow down
        subsequent computation;


COMMENT end of Fourier analysis example;

COMMENT the following program, written in  collaboration  with  David
Barton  and  John  Fitch,  solves a problem in general relativity. it
will compute the Einstein tensor from any given metric;

on nero;

COMMENT here we introduce the covariant and contravariant metrics;

operator p1,q1,x;

array gg(3,3),h(3,3);

gg(0,0):=e**(q1(x(1)))$
gg(1,1):=-e**(p1(x(1)))$
gg(2,2):=-x(1)**2$
gg(3,3):=-x(1)**2*sin(x(2))**2$

for i:=0:3 do h(i,i):=1/gg(i,i);

COMMENT generate Christoffel symbols and store in arrays cs1 and cs2;

array cs1(3,3,3),cs2(3,3,3);

for i:=0:3 do for j:=i:3 do
   begin for k:=0:3 do
        cs1(j,i,k) := cs1(i,j,k):=(df(gg(i,k),x(j))+df(gg(j,k),x(i))
                                     -df(gg(i,j),x(k)))/2;
        for k:=0:3 do cs2(j,i,k):= cs2(i,j,k) := for p := 0:3
                                 sum h(k,p)*cs1(i,j,p) end;

COMMENT now compute the Riemann tensor and store in r(i,j,k,l);

array r(3,3,3,3);

for i:=0:3 do for j:=i+1:3 do for k:=i:3 do
   for l:=k+1:if k=i then j else 3 do
      begin r(j,i,l,k) := r(i,j,k,l) := for q := 0:3
                sum gg(i,q)*(df(cs2(k,j,q),x(l))-df(cs2(j,l,q),x(k))
                + for p:=0:3 sum (cs2(p,l,q)*cs2(k,j,p)
                        -cs2(p,k,q)*cs2(l,j,p)));
        r(i,j,l,k) := -r(i,j,k,l);
        r(j,i,k,l) := -r(i,j,k,l);
        if i neq k or j>l
          then begin r(k,l,i,j) := r(l,k,j,i) := r(i,j,k,l);
                 r(l,k,i,j) := -r(i,j,k,l);
                 r(k,l,j,i) := -r(i,j,k,l) end end;

COMMENT now compute and print the Ricci tensor;

array ricci(3,3);

for i:=0:3 do for j:=0:3 do  
    write ricci(j,i) := ricci(i,j) := for p := 0:3 sum for q := 0:3
                                        sum h(p,q)*r(q,i,p,j);

COMMENT now compute and print the Ricci scalar;

rs := for i:= 0:3 sum for j:= 0:3 sum h(i,j)*ricci(i,j);

COMMENT finally compute and print the Einstein tensor;

array einstein(3,3);

for i:=0:3 do for j:=0:3 do
         write einstein(i,j):=ricci(i,j)-rs*gg(i,j)/2;

COMMENT end of Einstein tensor program;

clear gg,h,cs1,cs2,r,ricci,einstein;

COMMENT an example using the matrix facility;

matrix xx,yy,zz;

let xx= mat((a11,a12),(a21,a22)),
   yy= mat((y1),(y2));

2*det xx - 3*w;

zz:= xx**(-1)*yy;

1/xx**2;

COMMENT end of matrix examples;

COMMENT a physics example;

on div; COMMENT this gives us output in same form as Bjorken and Drell;

mass ki= 0, kf= 0, p1= m, pf= m;

vector eei,ef;

mshell ki,kf,p1,pf;

let p1.eei= 0, p1.ef= 0, p1.pf= m**2+ki.kf, p1.ki= m*k,p1.kf=
    m*kp, pf.eei= -kf.eei, pf.ef= ki.ef, pf.ki= m*kp, pf.kf=
    m*k, ki.eei= 0, ki.kf= m*(k-kp), kf.ef= 0, eei.eei= -1, ef.ef= -1; 

operator gp;

for all p let gp(p)= g(l,p)+m;

COMMENT this is just to save us a lot of writing;

gp(pf)*(g(l,ef,eei,ki)/(2*ki.p1) + g(l,eei,ef,kf)/(2*kf.p1))
  * gp(p1)*(g(l,ki,eei,ef)/(2*ki.p1) + g(l,kf,ef,eei)/(2*kf.p1))$

write "The Compton cross-section is ",!*ans;

COMMENT end of first physics example; 

off div;

COMMENT another physics example;

index ix,iy,iz;

mass p1=mm,p2=mm,p3= mm,p4= mm,k1=0;

mshell p1,p2,p3,p4,k1;

vector qi,q2;

factor mm,p1.p3;

order mm;

operator gga,ggb;

for all p let gga(p)=g(la,p)+mm, ggb(p)= g(lb,p)+mm; 

gga(-p2)*g(la,ix)*gga(-p4)*g(la,iy)* (ggb(p3)*g(lb,ix)*ggb(qi)*
    g(lb,iz)*ggb(p1)*g(lb,iy)*ggb(q2)*g(lb,iz)   +   ggb(p3)*
    g(lb,iz)*ggb(q2)*g(lb,ix)*ggb(p1)*g(lb,iz)*ggb(qi)*g(lb,iy))$

let qi=p1-k1, q2=p3+k1;

COMMENT it is usually faster to make such substitutions after all the
        trace algebra is done;

write "CXN =",!*ans;

COMMENT end of second physics example; 

showtime;
end;
\end{verbatim}    
\end{footnotesize}
\end{document}